%% file: NuclearAstrophysics.tex
\documentclass[12pt]{whitepaper}
\usepackage[colorlinks=true, linkcolor=blue, citecolor=blue, urlcolor=blue]{hyperref}
\usepackage{multicol}

\input{units}

\input{nuclides}

\newcommand{\ee}[1]{\ensuremath{\times 10^{#1}}}

\newcommand{\qhead}[1]{\vspace*{0.2cm} \noindent \textbf{\textsl{#1}}}
\newcommand{\cqc}{\href{http://www.nap.edu/openbook.php?record_id=10079&page=1}{\emph{Connecting Quarks with the Cosmos}}}
\newcommand{\frib}{\href{http://www.sc.doe.gov/np/program/FRIB.html}{FRIB}}
\newcommand{\dusel}{\href{http://www.int.washington.edu/DUSEL/}{DUSEL}}
\newcommand{\FeH}{\ensuremath{[\mathrm{Fe/H}]}}
\newcommand{\rsat}{\ensuremath{\rho_{0}}}

\newcounter{mypage}
\renewcommand*{\themypage}{\addtocounter{mypage}{1}\arabic{mypage}}
\newcommand{\pagelimit}{7}

\begin{document}
\ShortTitle{Nuclear Astrophysics}
\thispagestyle{empty}
\input{cover}
\clearpage
\setcounter{mypage}{0}
\cfoot{-\themypage/\pagelimit-}

\section{Introduction}

The field of nuclear astrophysics is enjoying a resurgence of activity. The nuclear science community has embraced many of the scientific questions of nuclear astrophysics. This white paper, directed to the Stars and Stellar Evolution panel, has three objectives: 1) to provide the \emph{Astro2010} Decadal Survey with a vista into the goals of the nuclear physics and nuclear astrophysics community; 2) to alert the astronomical community of joint opportunities for discoveries at the interface between nuclear physics and astronomy%
\footnote{%
The scope of this white paper concerns the specific area of ``low energy'' nuclear astrophysics. We define this as the area of overlap between astrophysics and the study of nuclear structure and reactions. The many other exciting scientific opportunities at the intersection of nuclear physics and astronomy, such as the science of neutrinos, cosmic rays, Dark Matter, or the state of matter in the early Big Bang, are hopefully being addressed by other white papers.};
and 3) to delineate efforts in nuclear physics and describe the observational and theoretical advances in astrophysics necessary to make progress towards answering the following questions in the 
\href{http://www.sc.doe.gov/np/nsac/nsac.html}{Nuclear Science 2007 Long Range Plan}.
\begin{tightitem}
\item \textbf{What is the origin and distribution of the elements?} 
\item{\textbf{What are the nuclear reactions that power stars and stellar explosions?}} 
\item{\textbf{What is the nature of dense matter?}} 
\end{tightitem}
Although the scope of this white paper is limited to these topics, we note that understanding the origin and distribution of the elements has an impact across wide swaths of astronomy. Two disparate examples are 1) the composition of pre-stellar gas clouds may affect the  likelihood of planet formation, retention of volatiles, and perhaps even the origin of life; and 2) accurate knowledge of CNO $\nu$-production rates would allow neutrino detectors to directly determine the metallicity of the solar core. Of the questions listed above, two---\emph{What is the origin of the elements?} and \emph{What is the nature of dense matter?}---were specifically listed in the National Academies Study \cqc.

\section{New Opportunities in Nuclear Physics}

Major experimental, observational, and theoretical advances in both nuclear physics and astronomy are required to address these scientific questions.  To address these needs, the nuclear physics community is making investments both in building new facilities and in building the collaborative infrastructure for an interdisciplinary endeavor; some examples are as follows.

\qhead{Facilities} The \href{http://www.sc.doe.gov/np/nsac/nsac.html}{Nuclear Science 2007 Long Range Plan} recommended the construction of the \$550M Facility for Rare Isotope Beams (\frib), previously endorsed by the NSAC and an independent National Academies Study (\href{http://www7.nationalacademies.org/bpa/RISAC_PREPUB.pdf}{RISAC report}). \frib\ will play a critical role in nuclear astrophysics by providing experimental access to most of the rare isotopes that govern stellar explosions and nucleosynthesis. 
Complementary to \frib\ are advances in nuclear astrophysics using beams of stable isotopes, and the expansion of efforts for deep underground measurements, such as the proposed Deep Underground Science and Engineering Laboratory (\dusel).

\qhead{Community} A prominent example is the NSF Physics Frontier Center JINA, (\href{http://www.jinaweb.org}{Joint Institute for Nuclear Astrophysics}). JINA enhances interdisciplinary efforts through virtual journals and seminars, organization of targeted workshops, and training young nuclear astrophysicists through participation in research, student exchange programs, and special schools. Similar efforts in Europe include the \href{http://www.gsi.de/forschung/helmholtz-alliance/EMMI.html}{ExtreMe Matter Institute} (EMMI) at \href{http://www.gsi.de}{GSI} (Darmstadt, Germany), the ``\href{http://www.universe-cluster.de/}{Origin and Structure of the Universe}'' research cluster (Munich, Germany) and the European network \href{http://www.ikp.physik.tu-darmstadt.de/carina/index.htm}{CARINA}. 

\section{Fundamental Questions of Nuclear Astrophysics}
\subsection{What is the origin and distribution of the elements?}

\emph{A fundamental challenge in nuclear astrophysics is our lack of a
 comprehensive understanding of the formation of trans-ferric elements}.
The site for the production of many of the elements heavier than iron, including gold, platinum, and uranium, is still unknown. It is thought that almost all of these isotopes are produced via the s(low)- and r(apid)- neutron-capture processes. Although the
formation of most of the heavier s-process elements can be accounted for by production
in low- to intermediate-mass AGB stars during their thermal pulsations 
\citep{Busso1999Nucleosynthesis},
there is no fully satisfactory astrophysical site established for the
origin of the r-process elements
\citep{Sneden2008Neutron-Capture}.

 Our understanding of the detailed processes that led to the production of the
elements is advanced greatly by the identification and spectroscopic analysis of
early-generation stars in the Galaxy. Astronomers can trace the origin of the s-process to a
``chemical time'' corresponding to $\FeH \sim -2.6$
\citep{Sneden2008Neutron-Capture}. The astrophysical origin of the r-process elements is best
probed by metal-poor stars that exhibit enhancements of these elements (relative
to solar ratios) by more than a factor of ten (the so-called r-II stars; see
\citep{Sneden2003The-Extremely-M}).
Whatever the astrophysical site of the
r-process is, recent observations suggest that it produced abundance patterns with remarkable
robustness over at least the first 8--10\nsp\Giga\yr\ of our Galaxy. It appears, however, that
an additional, new process of largely unknown nature contributes to the abundance of lighter "r-process" elements beyond iron.
Details remain murky due to the lack of an
adequately understood r-process site, incomplete knowledge of weak s-process contributions from carbon shell burning in massive stars, ignorance of the precise production ratios
of the actinides, and the observational fact that the abundances of the
actinides sometimes (in about 30\% of known cases) greatly exceeds their
expected level for low-metallicity stars with ages $> 10\nsp\Giga\yr$
\citep{Honda2004Spectroscopic-S}.

\qhead{Nuclear physics and astrophysical models}
Nuclear physics is needed to provide the data required to understand the underlying
reaction sequences. This must be coupled to advances in astrophysical modeling, including
multi-dimensional models of reactive flows, and multi-dimensional models of the convective cores of massive stars, and improvements in modeling core-collapse supernovae, to enable quantitative comparison between models and observations. Nuclear physics is also needed to
understand the nuclear reaction sequences that are most critical for the
synthesis of long-lived radioisotopes, which are important targets for observations. Significant progress has been made in
this area through many laboratory experiments, such as studies of the masses of
r-process waiting point nuclei \citep{Weber2009Nuclear-Masses-}.
Nevertheless, experimental studies of
r-process nuclei are still in their infancy, and the vast majority of the data
required to interpret the new observations have to wait for the planned
\frib.

\qhead{Galactic archaeology} By the time the dedicated stellar spectroscopic program
\href{http://www.sdss3.org/outermilkyway.php}{SEGUE-2} finishes (during
the first year of \href{http://www.sdss3.org/}{SDSS-III}), the SDSS efforts will have discovered a total of
some 25,000 stars with $\FeH < -2.0$, five times the number known just a few
years ago \citep{Beers2005The-Discovery-a}.
Several observational campaigns over the coming decade, including the
\href{http://www.lamost.org/en/}{LAMOST} survey in China,
\href{http://www.mso.anu.edu.au/~stefan/skymapper/index.php}{SkyMapper} in Australia,
and ESA's \href{http://www.rssd.esa.int/index.php?project=GAIA&page=index}{GAIA} satellite will provide a
wealth of chemical and dynamical information on these stars, including spectra
for huge numbers of stars in the Galactic disk and halo.  
Deeper surveys of the future, such as \href{http://www.lsst.org/lsst}{LSST}, 
will continue this upward trend to larger and larger
numbers of candidate low-metallicity stars, which will probe the chemical and dynamical
history of the Milky Way and its progenitor objects. The extreme paucity of stars with $\FeH < -3.5$ requires these large samples to ensure a decent sampling of the stars that have
recorded and preserved the Galactic chemical history.

\qhead{A comprehensive framework for chemical evolution}
A barrier to using the comprehensive databases coming online is a lack of a theoretical understanding to
maximally exploit this observational data. Efforts are being made to
develop a comprehensive framework for studying Galactic chemical evolution in a
cosmological context, and thereby follow the formation of the Galaxy
``one star at a time'' 
\citep{Tumlinson2006Chemical-Evolut}. This theoretical effort will be combined in the future with large-scale
simulations of galaxy formation and evolution to gain a detailed understanding
of the chemo-dynamical history of our Galaxy. At present, one of the fundamental
limitations to improvement of these models is the lack of reliable stellar evolution
models, which in turn are dependent on measured nuclear reaction rates for many
isotopes and nuclear processes. Over the lifetime of the Galaxy, stars sampling
nearly the entire range of the IMF contribute to the chemical composition of the
ISM. Thus, a detailed understanding of the recognized neutron-capture processes
is required to couple the nuclei produced by stars of various masses into the
chemical mix predicted for new generations of star formation.

\qhead{Follow-up spectroscopy} To glean clues about the detailed element production mechanisms requires high-resolution spectroscopic studies of the stars found by the surveys described above. 
Large-scale,
highly multiplexed high-resolution surveys, such as the
\href{http://www.sdss3.org/}{SDSS-III}
\href{http://www.sdss3.org/innermilkyway.php}{APOGEE} program, 
\href{http://www.aao.gov.au/AAO/HERMES/}{HERMES} at the AAT, and the proposed Gemini instrument
\href{http://www.gemini.edu/files/docman/science/aspen/WFMOS_feasibility_report_public.pdf}{WFMOS}
will provide the next great leap in our elemental abundance database for low-metallicity stars.   Although high-resolution
spectroscopic surveys are costly, they are necessary when studying the lowest-metallicity
stars ($\FeH\lesssim -3.0$), which have abundances that can only be determined from high-resolution (and high signal-to-noise) spectra.  
Extremely high signal-to-noise spectroscopy (using
a thirty meter-class telescope, for example) will enable the acquisition of isotopic
abundance information for a range of heavy elements, providing an even more finely-tuned
probe of the sites of the astrophysical r- and s-processs and of the ages of these stellar populations.

\qhead{Broader impacts} Detailed study of the shape of the metallicity distribution function of metal-poor 
halo stars, along with deviations, such as localized chemical peaks, and the distribution's low-metallicity terminus constrains the nature of the Galaxy's assemblage and the mass
function of the earliest sub-Galactic building blocks.
Furthermore, measurements of the observationally detectable
isotopic ratios in s-process-enhanced metal-poor stars
\citep{Aoki2003Europium-Isotop}
interfaces directly with nuclear physics models for their production.  
This work has a major
impact on studies at low-energy stable-beam accelerators, such as envisioned for \dusel.

\subsection{What are the nuclear reactions that power stars and stellar explosions?}\label{s.reactions}

\emph{Many of the most dramatic events in astronomy---X-ray bursts and superbursts, classical novae and white dwarf supernovae---are driven by nuclear reactions under explosive conditions. We understand poorly, however, the mechanics and reaction sequences occurring in these stellar explosions.} Our picture of X-ray bursts, thermonuclear explosions on the surfaces of neutron stars, has advanced greatly in the last decade.  Observations have, however, shown limits to our understanding and posed many new intriguing questions. 

As some examples, for X-ray bursts we do not fully understand the diversity of X-ray burst morphologies, including bursts much longer than the canonical 10--100\nsp\second\ durations; the occurrence of multiple bursts in a short time-span; the rare and energetic superbursts; and the order of magnitude discrepancy between the observed and predicted flux at which a transition to stable burning occurs. For classical novae, we still do not understand the cause of large CNO abundances in the ejecta and the dredge-up mechanism; the mass range of isotopes synthesized; or even whether there is a net gain or loss of mass over an accretion/explosion cycle. For Type Ia supernovae (SNe Ia), we do not understand which binaries evolve to produce SNe Ia, and are uncertain to what extent the explosion depends on the chemical and thermal properties of the progenitor white dwarf. There are large uncertainties on the nuclear physics side: for example, the cross section for $\carbon+\carbon$, the reaction that ignites SNe Ia and superbursts, is uncertain by several orders of magnitude at astrophysically relevant temperatures and densities \citep[see][]{Gasques2007Implications-of}. Even CNO reactions that power stars hold surprises: underground experiments are now reaching energies relevant for astrophysics and revising previous, extrapolated, cross-sections. These new measurements change, for example, the age estimate for the oldest globular clusters by $\lesssim 1\nsp\Giga\yr$ \citep{Imbriani2004The-bottleneck-}.

\qhead{Advances in nuclear experiment and theory}
The experimental determination of reaction rates on unstable nuclei that play a critical role in novae and X-ray bursts---in particular, the $\alpha$p- and rp-(rapid proton capture) processes---has just begun. Efforts to develop reaccelerated rare isotopes are underway to study the nuclei involved in these processes. Simulations indicate that SNe Ia are sensitive to the weak reaction rates used \citep{brachwitz.dean.ea:electron_captures}. If one is confident that the employed rates are reliable, one can use the nucleosynthesis patterns in the iron group to limit the central density and flame speed and, hence, validate explosion models. 
New experimental approaches, in which \frib\ (and other next-generation rare-isotope facilities in Europe, Japan, and Canada) will play a critical role, will expand studies of weak interactions to rare isotopes. These experimental approaches, working with
theoretical efforts, such as the  current \href{http://unedf.org/}{UNEDF} SciDAC program, will provide input for
neutrino scattering cross sections and electron capture rates, required
for supernovae modeling (both core-collapse and thermonuclear); binding energies, $\beta$-decay rates and particle capture rates, required for
the r-process and rp-process; and low-energy capture cross sections, required for
light nuclei reactions.
\emph{These programs, along with experimental studies of $\alpha p$- and rp-process nuclei, should be combined with sensitivity studies of nucleosynthesis calculations to delineate the accuracies required and to define targets for experimental study.}

\qhead{Defining the population}
Upcoming surveys with short cadences, such as \href{http://www.lsst.org/lsst}{LSST}, \href{http://pan-starrs.ifa.hawaii.edu/public/}{PanSTARRS}, \href{http://www.mso.anu.edu.au/~stefan/skymapper/index.php}{SkyMapper}, and the \href{http://www.astro.caltech.edu/ptf/}{Palomar Transient Factory}, are well-suited to discover rare and exotic explosions involving white dwarfs, such as He flashes from AM CVn stars \citep{Bildsten2007Faint-Thermonuc}. In addition to finding exotic new environments for explosive nucleosynthesis, such surveys will paint a more complete portrait of the accreting white dwarf family, which will help to constrain the progenitors of SNe Ia. 

\qhead{Spectroscopy}
During some X-ray bursts, the convective layer may extend far enough that material synthesized during the burst can be ejected during a radius expansion burst \citep{Weinberg2005Exposing-the-Nu}. The resulting edges are in principle detectable with Chandra or XMM, and will be within reach of missions such as the proposed  \href{http://ixo.gsfc.nasa.gov/}{International X-ray Observatory} (IXO).  In addition to providing a measurement of the gravitational redshift (see \S~\ref{s.ns}), such detections would give a unique measurement of the material synthesized in the burst, and hence, the composition of the outer crust of the neutron star. For classical novae, 
multi-wavelength spectroscopy of the ejecta could potentially reveal whether the nova experienced ``break out'' of the CNO cycle \citep{Glasner2009CNO-Breakout-an}. UV spectroscopy within the first few days of the outburst is sensitive to the atmospheric structure of the nova \citep[see][and references therein]{Gehrz1998Nucleosynthesis}. 
For SNe Ia, follow-up UV within the early phase of the lightcurve is sensitive to metallicity effects \citep{Lentz2000Metallicity-Eff}; X-ray spectroscopy of Ia supernovae remnants also shows promise for determining the progenitor metallicity \citep{Badenes2008The-End-of-Amne}.

\qhead{Nuclear decays: the fingerprints of the isotopes} The detection of $\gamma$-rays from decaying rare isotopes in space is a particularly sensitive probe of explosive nucleosynthesis, as well as of some extreme nucleosynthesis in 
stars. While elemental abundances typically can have many origins, 
including pre-existing material that has not undergone burning at all, 
radioactive isotopes are clearly produced in specific regions under 
extreme conditions. Detection of $\gamma$ lines is therefore a direct way to characterize the underlying processes in supernovae and novae. 
Transformational progress beyond the scope of the Compton Gamma Ray
Observatory and
\href{http://www.sciops.esa.int/index.php?project=INTEGRAL&page=index}{INTEGRAL}
missions (both have flux sensitities of $\sim 10^{-5} \nsp\gamma\usp\second^{-1}\usp\cm^{-2}$) requires 10--100 times better flux sensitivity. The technology to reach this flux sensitivity is under
development and includes the ACT mission concept, the GRASP mission
concept, and Laue diffraction lens.  With such improved flux
sensitivities, detection of 16--28 SNe Ia per year out to a distance of 100\nsp\Mega\parsec\ is feasible \citep{Timmes1997GammaaRay-Line-}.
Such detection capabilities would offer the possibility of detecting \titanium[44] in many supernova remnants, including1987A. Maps of the \iron[60] and \aluminum[26] distributions in the Galaxy 
could be compared to discover the sources of these radioactive isotopes. A first detection of \sodium[22] and 
\fluorine[18] in novae is also within reach and would be a major discovery 
providing unambiguous proof of the underlying thermonuclear processes. This would make it possible to probe the efficacy of convective mixing and to test assumptions about the underlying white 
dwarf mass. 
The detection of decay $\gamma$'s from rare isotopes 
ejected in X-ray bursts might be difficult even for an advanced mission, 
but a positive detection would offer a first 
direct tool to test the assumed nuclear processes powering these events.

\subsection{What is the nature of neutron stars and dense matter?}\label{s.ns}

\emph{We lack the ability to make quantitative predictions about nuclei from Quantum Chromodynamics (QCD).
Neutron stars are the unique  sites in the universe to study QCD at finite  
density, in precisely the same way that black hole horizons are  unique sites to study strong gravity.}
Neutron stars are among the most fascinating astrophysical objects: their structure and evolution are largely determined by nuclear physics, and they play a central role in many astrophysical events such as supernovae, X-ray bursts, and possibly gamma-ray bursts. Although born hot in the cauldron of a core-collapse supernova, a neutron star quickly cools, and its mechanical structure decouples from its temperature.  As a result, the basic observational properties of neutron stars---their masses and radii---are determined by the cold nuclear equation of state (EOS), which is a fundamental property of nuclear matter. 

\qhead{A unique laboratory for QCD}
Bulk matter at densities above the nuclear saturation density  
is routinely, if momentarily, produced in terrestrial colliders.  Under such conditions, the lab-frame energies of  
nucleons are far in excess of their rest mass, and the quarks which  
mediate interaction between the nucleons are effectively unbound.   
This is the realm of QCD at finite temperature.
Despite this progress, our knowledge of the properties of bulk nuclear matter is very incomplete.
Ongoing experiments with collisions of heavy nuclei such as tin can compress nuclear matter to several times its equilibrium density, and collisions with greater proton-neutron asymmetry will be possible with \frib. These experiments, however, cannot explore the low-temperature, high neutron-to-proton asymmetry regime. 
The only sites in the universe where one finds this matter are the cores of neutron stars. 

\qhead{Connecting Experiment and Observation} A critical component of the nuclear EOS is the symmetry energy, the amount of energy needed 
to increase the neutron-to-proton asymmetry. The density dependence of the symmetry energy 
controls a number of neutron star and nuclear matter properties: the neutron star mass-radius 
relation, the threshold density for the onset of rapid neutrino emission from the neutron star core, the 
thickness of the neutron ``skin'' in heavy nuclei, isovector excitations in nuclei (which depend on the 
difference between neutron and proton motions), isoscalar monopole experiments (which measure the nuclear compressibility using inelastic scattering), and the dynamics of intermediate energy heavy-ion 
collisions. Because this term in the EOS plays a role in many phenomena, there is the exciting 
possibility of looking for correlations between data from current and planned nuclear experiments and data from observations over the electromagnetic, and possibly gravitational, spectrum. Gravitational wave detectors already have the sensitivity to detect the emission from ``mountains'' on neutron stars with exotic equations of state \citep{Owen2005Maximum-Elastic}, and within the next decade advanced detectors could begin placing limits on the crust deformation for ``canonical'' neutron stars \citep{ushomirsky.cutler.ea:deformations}.  This is directly affected by the thickness and mechanical properties of the crust.  Nuclear experiments are probing the EOS at these densities. Realistic crust compositions, are being computed by following the ``ashes'' of X-ray bursts (see \S~\ref{s.reactions}) \citep{Gupta2008Neutron-Reactio}, and experiments on neutron-rich nuclei with \frib\ will inform our knowledge of the crust composition. This knowledge is a necessary input for interpreting potential detections of gravitational wave emission from deformed neutron stars.

A variety of experiments have constrained the EOS of symmetric nuclear matter at densities 0.4--4\usp\rsat, where  $\rsat \approx 2.6\ee{14}\nsp\grampercc$ is the nuclear saturation density (the density of matter within a nucleus). The EOS can link laboratory measurements to the properties of neutron stars in intriguing ways. For example, measurements of the thickness of the neutron skin of a heavy nucleus, or the flow of neutrons and protons in nuclear collisions can be linked to measurements of the radii of 
neutron stars with known masses. Additional correlations have emerged between the neutron skin 
thickness of a heavy nucleus, the flow of neutrons and protons or the production of positive and 
negative pions in nuclear collisions to quantities such as neutrino cooling rates, the stellar moments 
of inertia, or vibration frequencies of the crust (the outer region of the neutron star 
where individual nuclei exist). Laboratory measurements and neutron star observations display 
different and therefore complementary sensitivities to the EOS. Studies of 
collisions and neutron stars probe very different regions and possibly different phases in the QCD 
phase diagram for $\rho\lesssim 10\usp\rsat$. \emph{Within the next decade, there is a real possibility of combining laboratory experiment and astronomical observation to determine the cold equation of state for bulk nuclear matter for densities up to ten times the saturation density.}

\qhead{The neutron star mass-radius relation} 
There is a wide assortment of neutron star observations that can in principle inform us about the nature of dense matter \citep[see][for an overview]{Lattimer2006Neutron-Star-Ob}, and we give here a few examples: timing measurements of binary pulsars determine mass $M$; spectroscopy of lines determines $1+z = (1-2GM/Rc^{2})^{-1/2}$; continuum spectral fitting of thermal emission determines radii $R^{\infty}/d = R(1+z)/d$, where $d$ is the distance; and cooling of isolated neutron stars and neutron star transients determines the strength of the neutrino emissivity.

In general, massive neutron stars are not accomodated by families of equations of state with substantial softening at high densities. A single, well-constrained, mass measurement greater than $\approx 2\nsp\Msun$ would spur enormous theoretical efforts, for example.  \emph{For determining the EOS, however, what is needed ideally are several neutron stars with each having a well-measured $M$ and $R$}.  Because the dense matter EOS maps to a unique mass-radius relation \citep{Lindblom1992Determining-the}, measurements of several ($M, R$) pairs can in principle be used to empirically determine the equation of state \citep{Postnikov2009Inverting-the-d}. High-quality spectra during X-ray bursts can in principle provide independent measurements of quantities, such as the Eddington flux, that depend on both $M$ and $R$, so that a joint constraint is possible \citep{Ozel2006Soft-Equations-}. Continuum fits to the thermal spectra of quiescent neutron stars can also simultaneously constrain $M$ and $R$ \citep{Heinke2006A-Hydrogen-Atmo}.  This will require 0.1\% accuracy per energy channel, over $0.5\textrm{--}2\nsp\keV$, improvements in the computational precision of spectral models, and distance measurements that are accurate to 3\% out to 5\nsp\kilo\parsec.

\qhead{Finding rare events}
A variety of phenomena depend on the neutrino emissivity, thermal conductivity, shear viscosity and viscosity of the neutron star: cooling of isolated neutron stars \citep{yakovlev.pethick:neutron} and quasi-persistent transients \citep{Cackett2008Cooling-of-the-}, energetics and recurrence times of superbursts and long X-ray bursts\citep{Cumming2005Long-Type-I-X-r}, and vibration frequencies during magnetar flares \citep{Strohmayer2006The-2004-Hyperf}, to give a few examples. Many of these depend on the detection and follow-up of rare events.  It is critically important that there be a capability to continuously monitor X-ray sources and to respond quickly to targets of opportunity. 

\begin{multicols}{2}
\footnotesize
\bibliographystyle{apj}
\bibliography{nucastro}
\end{multicols}
\end{document}

%% file: units.tex
\newcommand{\unitspace}{\ensuremath{\,}}
\newcommand{\usp}{\unitspace}
\newcommand{\numberspace}{\ensuremath{\;}}
\newcommand{\nsp}{\numberspace}
\newcommand{\unitstyle}[1]{\ensuremath{\mathrm{#1}}}
\newcommand{\power}[2]{\ensuremath{{#1}^{#2}}}

\newcommand{\centi}{\unitstyle{c}}
\newcommand{\kilo}{\unitstyle{k}}
\newcommand{\Mega}{\unitstyle{M}}
\newcommand{\Giga}{\unitstyle{G}}

\newcommand{\meter}{\unitstyle{m}}

\newcommand{\second}{\unitstyle{s}}

\newcommand{\cm}{\centi\meter}
\newcommand{\gram}{\unitstyle{g}}

\newcommand{\grampercc}{\gram\usp\power{\cm}{-3}} 

\newcommand{\eV}{\unitstyle{eV}}        
\newcommand{\keV}{\kilo\eV} 

\newcommand{\Msun}{\ensuremath{M_\odot}}

\newcommand{\parsec}{\unitstyle{pc}}

\newcommand{\yr}{\unitstyle{yr}}        


%% file: nuclides.tex
\newcommand{\nuclei}[2]{\ensuremath{\mathrm{^{#1}#2}}}

\newcommand{\helium}[1][4]{\nuclei{#1}{He}}

\newcommand{\carbon}[1][12]{\nuclei{#1}{C}}

\newcommand{\fluorine}[1][19]{\nuclei{#1}{F}}

\newcommand{\sodium}[1][23]{\nuclei{#1}{Na}}

\newcommand{\aluminum}[1][27]{\nuclei{#1}{Al}}

\newcommand{\calcium}[1][40]{\nuclei{#1}{Ca}}

\newcommand{\titanium}[1][48]{\nuclei{#1}{Ti}}

\newcommand{\chromium}[1][52]{\nuclei{#1}{Cr}}

\newcommand{\iron}[1][56]{\nuclei{#1}{Fe}}

\newcommand{\nickel}[1][58]{\nuclei{#1}{Ni}}

\newcommand{\tin}[1][120]{\nuclei{#1}{Sn}}

\newcommand{\lead}[1][208]{\nuclei{#1}{Pb}}

%% file: cover.tex
\begin{center}
\Huge \bfseries\sffamily Nuclei in the Cosmos \\ 
\vspace{0.1in}
\Large A White Paper submitted to the \emph{Astro2010}\\ Decadal Survey of Astronomy and Astrophysics\\
\vspace{0.2in}

\normalsize\mdseries
\begin{tabular}{ll}
Primary Panel & \textsl{Stars and Stellar Evolution (SSE)}\\
Secondary Panels & \textsl{The Galactic Neighborhood (GAN)}\\
   &  \textsl{Galaxies across Cosmic Time (GCT)}\\
\end{tabular}
\vspace{0.8in}

\sffamily Edward Brown\\
Department of Physics and Astronomy \\
Michigan State University	\\
\href{mailto:ebrown@pa.msu.edu}{ebrown@pa.msu.edu}
\vspace{0.15in}

\sffamily
\begin{tabular}{lr}
\input{authors}
\end{tabular}

\end{center}

%% file: authors.tex
Timothy C. Beers    & Michigan State University         \\
B. Alex Brown    & Michigan State University     \\
Carl Brune          & Ohio University                   \\
Art Champagne       & University of North Carolina      \\
Christian Illiadis  & University of North Carolina      \\
William Lynch       & Michigan State University         \\
Brian O'Shea        & Michigan State University         \\
Peter Parker         & Yale University                   \\
Robert Rutledge     & McGill University                 \\
Michael Smith       & Oak Ridge National Laboratory     \\
Sumner Starrfield   & Arizona State University          \\
Andrew Steiner      & Michigan State University         \\
Frank Timmes        & Arizona State University          \\
James Truran        & University of Chicago           \\
Michael Wiescher    & University of Notre Dame          \\
Remco Zegers        & Michigan State University         \\